\documentclass[aip,apl,amsmath,amssymb,reprint,numerical,graphicx]{revtex4-1}
\usepackage{graphicx}
\usepackage{dcolumn}
\usepackage{bm}
\draft 

\begin{document}


 \title{Charge and spin Hall effect in spin chiral ferromagnetic graphene}

\author{Babak Zare Rameshti} 
\author{Malek Zareyan}
\email{zareyan@iasbs.ac.ir}
\affiliation{$^{1}$Department of Physics, Institute for Advanced Studies
in Basic Sciences (IASBS), Zanjan 45137-66731, Iran }

\date{\today}

\begin{abstract}
We predict a specific type of {\it charge} Hall effect in undoped ferromagnetic graphene that is generated by the {\it spin} Hall mechanism in  the absence of an external magnetic field.  The essential feature is the so-called spin chiral configuration of the spin subbands in such a magnetic material where carriers with opposite spin direction are of different type of electron-like or hole-like. Within the semiclassical theory of spin-orbital dynamics of electrons, we obtain that a longitudinal electric field can produce a spin-orbit  transverse current of pure charge with no polarization of the spin and the valley.  
\end{abstract}

\maketitle

The Hall effect refers to the generation of a voltage difference across a conductor when an electrical current flowing along the conductor is subjected to a perpendicular magnetic field \cite{Hall1879}.  This effect can be well explained within the classical Drude model \cite{Drude1900}, which describes how the charges of opposite sign accumulate on the opposing edges of the conductor to compensate for the Lorentz force diverting  electrons in the transverse direction. The spin analogues of Hall effect is an intriguing phenomena, introduced first by  Dyakonov and Perel \cite{DP1971}, where a charge current induces spin accumulation of opposite polarization on the lateral edges of a conductor. The intriguing feature of the spin Hall effect is that no external magnetic field is needed for its realization. Instead, it is caused by the spin-orbit interaction which couples spin and charge currents and leads to the spatial separation of carriers with different spin direction \cite{Hirsch99,Chud07}. The inverse of the spin Hall effect is also possible, which converts the spin-current into a transverse charge current through the spin-orbit potential. Both direct and inverse effects were demonstrated experimentally in semiconductors \cite{Kato,Wunderlich,Bakun,Zhao}, as well as in metals  \cite{Saitoh,Valen,Kimura}.

\begin{figure}[ht!]
\includegraphics[trim=5cm 12.25cm 7.75cm 4.5cm, clip, scale=0.75]{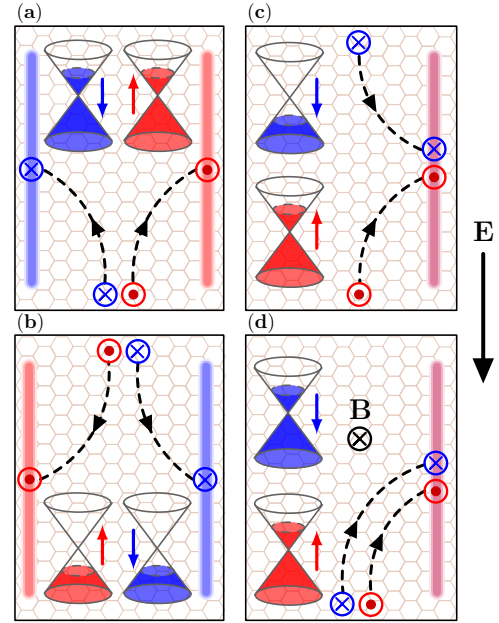}
\caption{(color online) Qualitative derivation of the predicted charge spin-Hall effect and its comparison to the ordinary Hall effect.  The spin Hall effect is represented in normal graphene doped with electrons (a) and holes (b), and in an undoped ferromagnetic graphene (c ) where carriers with up spin and down spin are of electron- and hole-like, respectively. The configuration of the two spin sub bands (being electron-like or hole-like) together with schematic of their spatial separation by the spin-orbit force are shown.  From a combination of the spin Hall effect for spin up electrons in (a) and spin down holes in (b) the effect can be constructed in (c ), which shows that  in ferromagnetic graphene carriers with different spin states acquire transverse drift velocities of the same direction. This creates a Hall current of pure charge. (d) Ordinary charge Hall effect in an electron-doped normal graphene where the carriers are diverted by the Lorentz force of an external magnetic field ${\bf B}$. The same electric field ${\bf E}$ is applied  in all four cases.}\label{fig1}
\end{figure} 

In this Letter, we report on the possibility of a different type of Hall effect which despite generating a transverse current of pure charge, is originated from the spin Hall mechanism in the absence of an external magnetic field. Our studied effect is based on pristine ferromagnetic graphene in which the sub bands of charge carriers with opposite spin direction are of different types of  electron(n)-like or hole(p )-like. This so called spin chiral configuration of the Dirac spin subbabds \cite{ZareyanSpinChiral1,ZareyanSpinChiral2,ZareyanSpinChiral3} has been recently realized in the experiment \cite{GeimGiantNonlocality} through a giant nonlocal effect in magneto-electronic response of undoped graphene  \cite{GiantSH,Nikolic12} . Here, we show that in such a system the spin-orbit interaction can cause diversion of electrons with opposite spin in the {\it same} transverse direction, which in combination  with the property of electron-hole symmetry of Dirac subbands in graphene, result in a net transverse charge current with {\it no} spin polarization. Within a semiclassical theory for spin-orbital dynamics of electrons in a ferromagnetic graphene, we find that the amplitude of  this Hall effect is determined by the strength of the spin-orbit coupling and the exchange energy. 

We start with a qualitative derivation of this charge Hall effect by analyzing the polarization of the transverse spin current in the spin Hall state of a normal graphene when its type of doping is changed from n to p.  The key feature is that in the diffusive limit an externally applied electric field drives a longitudinal drift velocity that for p carriers is oriented oppositely to that of n carriers (see Fig. \ref{fig1} a and b).  As we will see in the following this follows from the fact that a hole-like carrier in the valance Dirac band moves opposite to its wave-vector in contrast to an electron-like carrier of the conduction Dirac band which moves along its wave vector. The switching of the direction of  the drift velocity by changing the carrier type together with the dependence of the spin-orbit interaction on  the spin-direction and the momentum leads us to the conclusion that  p-type carriers with an specific spin direction acquires a transverse drift velocity which is opposite to that of  n-type carriers with the same spin direction. Consequently, polarizations of the spin Hall current are oppositely oriented in n-doped  and p-doped graphene.  Now we can construct the spin-resolved configuration of the drift velocity in the spin-chiral ferromagnetic graphene by  a combination of the spin Hall effect for up spin electrons in n-doped case (Fig. \ref{fig1}a) and down spin holes in p-doped case (Fig. \ref{fig1}b). From this observation we obtain that in the ferromagnetic graphene the carriers of different spin directions acquire drift velocities in the same transverse direction when diverted by the spin-orbit force (see Fig. \ref{fig1} c). This concludes our qualitative derivation that in such a system the spin Hall effect results in a transverse current of pure charge with zero polarization of the spin. It is worth noting the difference between this Hall effect and the ordinary Hall effect which is induced by an external magnet field (see Fig. \ref{fig1} d). 
We also note the resemblance between our predicted charge Hall effect and the inverse spin Hall effect in ordinary two dimensional electronic systems where the spin-orbit coupling of an applied spin current produces  a transverse charge current. The difference is that in our effect the spin current is driven by applying of an electric field to the spin chiral subbands, as described above.   

Before developing a quantitative theory we briefly comment on the experimental realization of ferromagnetic state in graphene. In addition to the possibility of an intrinsic ferromagnetic ordering in graphene nanoribons \cite{RibonFerro}, progress have already been made in  proximity-induced spin polarization by fabrication of transparent contacts between a graphene monolayer and ferromagnetic electrodes  (see for instance, Refs. \cite{grapheneFerroMetal1,grapheneFerroMetal2,grapheneFerroMetal3}). More recently, induction of a large exchange splitting has been demonstrated by depositing ferromagnetic insulator EuO on graphene\cite{InsulFerro}. 

To develop the semiclassical theory of the spin Hall effect \cite{Chud07} in a ferromagnetic graphene, we  consider the following Hamiltonian in the spin-${\bf \sigma}$ space
\begin{equation}
H=\alpha v_{F}k+E_{F}+\sigma_{z}h +V({\bf r})+\lambda \left[{\bf \sigma}\times \nabla_{\bf r}V({\bf r})\right]\cdot {\bf k}\label{a1}
\end{equation}
in which ${\bf \sigma}\equiv (\sigma_x,\sigma_y,\sigma_z)$ is the vector of Pauli matrices and $\alpha$ is equal to the unit matrix in the spin space for the exchange energy $h$ smaller than the Fermi energy $E_F$ and equal to $\sigma_{z}$ for $h>E_F$. This takes into account the fact that while for $h<E_F$ both of the Dirac spin sub bands are of n-type, for $h>E_F$ the sub band of spin-up electrons becomes of p-type with its Fermi level being shifted to the valance subband. Here $V({\bf r})$ represents the total potential which include effects of periodic crystal potential $V_{c}$,  potential due to static disorders of the crystal lattice $V_{i}$, and an external potential $V_{e}$. The spin orbit interaction is taken into account by the last term in Eq.\ (\ref{a1}) with  an strength $\lambda$.
\par
From the Hamiltonian (\ref{a1}) we obtain semiclassical equation of motion for canonically conjugated position ${\bf r}$ and wave vector ${\bf k}$ as
\begin{eqnarray}
\dot{{\bf r}}&=&\nabla_{\bf k} {H}=\alpha v_{F}\hat{{\bf k}}+\lambda\left({\bf \sigma}\times \nabla_{\bf r}V\right)\label{a3}
\\
\dot{{\bf k}}&=&-\nabla_{\bf r}{H}=-\nabla_{\bf r} V-\lambda \nabla_{\bf r} \left[\left({\bf \sigma}\times \nabla_{\bf r}V\right)\cdot{\bf k} \right]\label{a4}
\end{eqnarray}
where $\hat{{\bf k}}$ is the unit vector along the wave vector ${\bf k}=k\hat{{\bf k}}$. From time derivative of Eq. (\ref{a3}) one obtains
\begin{equation}
\alpha v_{F}\dot{\hat{{\bf k}}}=\ddot{{\bf r}}-\lambda\left({\bf \dot{r}}
\cdot \nabla_{\bf r} \right)\left({\bf \sigma}\times \nabla_{\bf r}V\right),\label{a8}
\end{equation}
which in combination with Eqs. (\ref{a3}) and (\ref{a4}) results in the following 
form of the equation of motion for charge carriers
\begin{equation}
\ddot{{\bf r}}=-\alpha \frac{v_{F}}{k}(\nabla_{\bf r}V +{\dot k} \hat{{\bf k}})-\lambda\dot{{\bf r}}\times\left[\nabla_{\bf r}\times\left({\bf \sigma}\times\nabla_{\bf r}V\right)\right] \label{a13}
\end{equation}
In obtaining the above result, we have neglected the term proportional to $\lambda^2$. To take into account the scattering of the carriers from static disorders, we follow the Drude model by adding to Eq.\ (\ref{a13}) the damping term $-\dot{\bf{r}}/\tau$, with $\tau$ being the corresponding mean free time. In this way we obtain the final form of the equation of motion as
\begin{equation}
\ddot{{\bf r}}=-\alpha \frac{v_{F}}{k}(\nabla_{\bf r}V +{\dot k} \hat{{\bf k}})-\lambda\dot{{\bf r}}\times\left[\nabla_{\bf r}\times\left({\bf \sigma}\times\nabla_{\bf r}V\right)\right] -\frac{\dot{{\bf r}}}{\tau}, \label{a14}
\end{equation}
with $V$ in Eq.\ (\ref{a14}) is now excluding the effect of the 
disorder potential.  We have solved Eq.\ (\ref{a14}) by linearizing it with 
respect to $e{\bf E}=-\nabla_{\bf r}V_e$ and keeping the terms up to order  $\lambda$. 
After a volume averaging , the solutions take the form
\begin{eqnarray}
\langle \dot{\bf r}_{\sigma_{z}} \rangle  &=& \alpha \frac{e\tau v_{F}}{k}\left[
{\bf E} -\tau \lambda {\bf E}\times\left\langle
\nabla_{\bf r} \times\left({\bf \sigma}\times
\nabla_{\bf r} V_c \right) \right\rangle \right],\qquad
\label{r1}
\end{eqnarray}
The right hand side of Eq.\ (\ref{r1}) contains the volume average
of $\nabla_{{\bf r}i}\nabla_{{\bf r}j}V_c({\bf r})$, which for the two dimensional hexagonal lattice of graphene
 takes the form $\left\langle \nabla_{{\bf r}i}\nabla_{{\bf r}j}V_c\right\rangle = C \delta_{ij}$,
where $C$ is a constant to be determined from the atomic parameters \cite{Chud09}
\par
From Eq.\ (\ref{r1})
one obtains the transverse component of the mean drift velocity as:
\begin{equation}\label{spin-velocity}
\langle \dot{\bf r}_{ \sigma_{z}\perp} \rangle = \frac{\lambda C e^{2}\tau^{2}v_{F}}{k}\alpha ({\bf \sigma}\times {\bf E})\,.
\end{equation}
The current density carried by spin-${\sigma_{z}}$ electrons is calculated by the relation
 \begin{equation}
{\bf j}_{\sigma_{z}} = e\int \frac{d^2{\bf k}}{(2 \pi)^2} \langle \dot{\bf r}_{\sigma_{z}}\rangle,
\label{jsigmaz}
\end{equation}
where the integration is carried out over the occupied ${\bf k}$-states by taking into account the spin-dependent electronic density of states  $N_{\sigma_{z}}(\varepsilon)=(N_{F}/2)\left\vert 1+\varepsilon/E_F+\sigma_{z}h/E_F\right\vert$ at an excitation energy $\varepsilon$; $N_{F}=k_F/(\pi v_F) $.  Using Eqs. (\ref{r1}), (\ref{spin-velocity}), and (\ref{jsigmaz}) we obtain
${\bf j}_{\sigma_{z}}$ at zero temperature, through which we can calculate the charge current density as
\begin{eqnarray}
{\bf j}_{c}=\sum_{\sigma_{z}=\pm 1}{\bf j}_{\sigma_{z}}
=\sigma_D[ {\bf E}+e\tau \lambda C \frac{h}{E_F} ({\bf \hat{z}}\times{\bf E}) ].
 \label{a25}
\end{eqnarray}
Similarly for the spin current density, we obtain
\begin{eqnarray}
{\bf j}_{s}=\sum_{\sigma_{z}=\pm 1}\sigma_{z}{\bf j}_{\sigma_{z}}
=\sigma_D[ \frac{h}{E_F} {\bf E}+e\tau \lambda C ({\bf \hat{z}}\times{\bf E})]\label{a26}.
\end{eqnarray}
where we have taken the polarization of the spin ${\bf \sigma}=\sigma_z {\bf \hat{z}}$ to be along the $z$-axis. Here $\sigma_{D}=e^{2}\tau E_F/\pi$ is the normal state Drude conductivity of graphene.

Let us start with analyzing the results (\ref{a25}) and  (\ref{a26}) for undoped ferromagnetic graphene $E_F=0$. In this case we obtain:
\begin{eqnarray}
{\bf j}_{c}&=& \frac{e^3\tau^2 \lambda C}{\pi} h ({\bf \hat{z}}\times{\bf E}),
 \label{a25b}\\
{\bf j}_{s}& =&\frac{e^2\tau}{\pi} h {\bf E}\label{a26b},
\end{eqnarray}
which show that the transverse current contains only a charge component. This implies that the applied electric field creates through the spin-orbit potential a transverse charge current without spin polarization. Interestingly, in this limit the electric field can drive only spin current in the longitudinal direction and there is no net charge flow along the field. We note that the amplitude of this spin Hall mechanism induced charge Hall conductance is linearly proportional to the exchange splitting $h$.  Therefore, even for a weak spin-orbit interaction, a large transverse conductance can be achieved by inducing a large spin splittings \cite{InsulFerro}. Moreover, creating a large spin-orbit coupling in graphene by using heavy addatoms \cite{GrapheneSH01} or substrate impurities \cite{GrapheneSH02,GrapheneSH03} is currently of high interest to demonstrate experimentally the related intriguing phenomena \cite{GrapheneSH1,GrapheneSH2,GrapheneSH3} including the quantum spin Hall effect \cite{KaneMele}. Thus the experimental observation of our studied charge spin-Hall effect is expected to be feasible.

The absence of spin polarization in the Hall current is the result of the fact that the hole-like  spin-down carriers acquires a transverse drift velocity which has the same direction as that of electron-like  spin-up carriers (see Eq. (\ref{spin-velocity})). Furthermore, for $E_F=0$ there is an electron-hole symmetry such that the Fermi level density of states for spin-up and spin-down carriers are the same. Thus, the carriers with opposite spin generate equal current densities in the same transverse direction.   A finite doping changes the above picture by producing a difference in the Fermi level density of states of spin-up electrons and spin-down holes. This in turn will produce a finite transverse spin current as well as a finite longitudinal charge current, both being proportional to the doping $E_F$. In the limit of high dopings $E_F\gg h$, the ordinary spin Hall effect is recovered with a longitudinal charge current density ${\bf j}_{c}=\sigma_D {\bf E}$ and a transverse spin current density ${\bf j}_{s}=\sigma_De\tau \lambda C ({\bf \hat{z}}\times{\bf E})$.

We note that up to now the valley degree of freedom of electrons has been disregarded. In fact our results are valid for one of the two valleys, say the valley $K$.  We can extend our calculations to electrons in the other valley $K^{\prime}$, by considering the Hamiltonian for this valley. This can be constructed by making the time-reversal transformation of the Hamiltonian (\ref{a1}) of the valley $K$, which from invariance of the spin-orbit potential under this transformation, is turned out to be the same as the Hamiltonian (\ref{a1}).
Thus our results does not depend on the valley degree of freedom, namely, the carriers of the valley $K^{\prime}$ contribute the same charge and spin current densities as those of the valley $K$ (Eqs. (\ref{a25b}) and (\ref{a26b})). This leads us to the conclusion that ferromagnetic graphene with zero doping to exhibit a transverse charge conductance $\sigma_c=2e^{3}\tau^2 \lambda C h/\pi$ and a longitudinal spin conductance $\sigma_{s}=2e^{2}\tau h/\pi$ with a ratio $\sigma_{c}/\sigma_{s}=e\tau \lambda C$.  We further conclude that this charge spin-Hall effect does not exhibit {\it valley} polarization.  

It is also interesting to note that there should be an inverse of this charge Hall effect since it is originated from the spin Hall process, the process whose inverse exists. In the direct effect at $E_F=0$, an applied electric field generates a pure longitudinal spin current whose amplitude can be controlled by the exchange splitting. This generation of the spin current, which itself can also be of great interest in the field of spintronics \cite{SpintronicsRev1,SpintronicsRev2}, is a peculiar one: it occurs in the condition of an applied longitudinal {\it charge} potential and without induction of any local spin accumulations. Denoting the spin-resolved longitudinal voltage difference by $V_{\sigma_z}$, thus, we have  $V_c=V_{+}+V_{-} \neq 0$ and $V_s=V_{+}-V_{-}=0 (V_{+}=V_{-})$ despite having no longitudinal charge current but a finite spin current. The spin-orbit induced transverse charge current also exhibits a quite unusual property as it can create a lateral accumulation of the {\it spin} and an associated spin potential over the edges $V_{\perp s}=V_{\perp +}-V_{\perp -}\neq 0$ while  the lateral charge potential $V_{\perp c}=V_{\perp +}+V_{\perp -}= 0 (V_{\perp +}=-V_{\perp -})$, with $V_{\perp \sigma_{z}}$ being the transverse voltage difference for spin $\sigma_z$ carriers.  Now, the inverse effect can be generated by a longitudinal current of pure charge in the presence of a spin voltage. A transverse current of pure spin is then induced by the spin-orbit coupling, which can be detected as a lateral charge potential. To summarize, the direct effect generates a transverse charge current and a lateral spin voltage from an applied longitudinal charge voltage while the inverse effect generates a transverse spin current and a lateral charge potential from a longitudinal charge current.  These unusual features are the results of the spin-chiral configuration of the Dirac cone spin sub bands, as explained above (see Fig. \ref{fig1} c).

Finally, It is important to note that our predicted Hall effect is fundamentally different from the anomalous Hall effect (AHE) in ordinary ferromagnetic metals\cite{AHERev1,AHERev2}. In AHE the spin-orbit induced transverse Hall current, either having an intrinsic Berry phase \cite{AHEIn} or an extrinsic spin-dependent impurity scattering \cite{AHEEx} origin, is carried by carriers with an imbalanced spin states of single-type (either electron-like or hole-like). While our predicted effect is based on a connection between the spin state and the type of  carriers which can be realized by spin-chiral ferromagnetic graphene. This difference can be also manifested by response of spin-chiral bands to 
an external magnetic field $B$, which we have obtained to extend the results (\ref{a25b},\ref{a26b}) to       
\begin{eqnarray}
{\bf j}_{c}&=& \frac{e^3\tau^2 \lambda C}{\pi} h ({\bf \hat{z}}\times{\bf E}),
 \label{a25bb}\\
{\bf j}_{s}& =&\frac{e^2\tau}{\pi} h {\bf E}-\frac{e^3\tau^2 B}{\pi} h ({\bf \hat{z}}\times{\bf E}).
\label{a26bb},
\end{eqnarray}
These results shows that external magnetic field induces only a pure spin current in the transverse direction even in the absence of an spin-orbit interaction. This situation is in striking contrast to AHE.

In conclusion, we have demonstrated realization of a special type of magneto electronic transport in a conducting material whose charge carriers with up and down spin direction are of electron- and hole-like, respectively. Such a material has been recently realized experimentally  \cite{GeimGiantNonlocality} in undoped magnetic graphene with Zeeman splitted spin subbands of Dirac type. We have found that an external electric field  can drive a pure spin current through this material with a spin conductivity proportional to the splitting energy. The spin-orbit coupling of this spin-current creates a transverse current of pure charge, which presents a specific charge Hall effect with no need of an external magnetic field. We have also explored the inverse of this charge Hall effect in which a pure longitudinal charge current generates a transverse spin current with an associated charge Hall voltage. Our study reveals the suitability of magnetized graphene for generating spin current, that is of crucial importance for applications in spintronic technology.


We acknowledge fruitful discussions with S. Abedinpour and A. G. Moghaddam. M. Z. thanks G. E. W. Bauer of the Institute for Materials Research of Tohoku University in Sendai, and the Associate Office of ICTP in Trieste for the hospitality and support during his visit to these institutes.


\begin{thebibliography}{99}
\bibitem{Hall1879}
E. Hall, American Journal of Mathematics {\bf 2}, 287 (1879).

\bibitem{Drude1900}
P. Drude, Ann. Phys. {\bf 1}, 566 (1900); P. Drude, Ann. Phys. {\bf 3}, 369 (1900).

\bibitem{DP1971}
M. I. Dyakonov and V. I. Perel, Sov. Phys. JETP Lett. {\bf 13}, 467 (1971); 
M. I. Dyakonov and V. I. Perel, Phys. Lett. A  {\bf 35}, 459 (1971).

\bibitem{Hirsch99}
J. E. Hirsch, Phys. Rev.  Lett. {\bf 83}, 1834 (1999).

\bibitem{Chud07}
E. M. Chudnovsky, Phys. Rev. Lett. \textbf{99}, 206601 (2007).

\bibitem{Kato}
Y. K. Kato, R. C. Myers, A. C. Gossard, and D. D. Awschalom,
Science {\bf 306}, 1910 (2004).

\bibitem{Wunderlich}
J. Wunderlich, B. Kaestner, J. Sinova, T. Jungwirth, Phys. Rev.
Lett. {\bf 94}, 047204 (2005).

\bibitem{Bakun}
A. A. Bakun, B. P. Zakharchenya, A. A. Rogachev, M. N. Tkachuk, and V. G. Fleisher, Sov. Phys. JETP Lett. {\bf 40}, 1293 (1984).

\bibitem{Zhao}
H. Zhao, E. J. Loren, H. M. van Driel, and A. L. Smirl, Phys. Rev. Lett. \textbf{96}, 246601 (2006).

\bibitem{Saitoh}
E. Saitoh, M Ueda, H. Miyajima, and G. Tatara, Appl. Phys. Lett \textbf{88}, 182509 (2006).

\bibitem{Valen}
S.O. Valenzuela, M. Tinkham, Nature \textbf{442}, 176 (2006).

\bibitem{Kimura}
T. Kimura, Y. Otani, T. Sato, S. Takahashi, and S. Maekawa, Phys. Rev. Lett. \textbf{98}, 156601 (2007).

\bibitem{ZareyanSpinChiral1}
M. Zareyan, H. Mohammadpour, and A. G. Moghaddam, Phys. Rev. B {\bf 78}, 193406 (2008) 

\bibitem{ZareyanSpinChiral2}
A. G. Moghaddam and M. Zareyan Phys. Rev. Lett. {\bf 105}, 146803 (2010).

\bibitem{ZareyanSpinChiral3}
M. V. Hosseini and M. Zareyan,  Appl. Phys. Lett \textbf{101}, 252602 (2012).

\bibitem{GeimGiantNonlocality}
D. A. Abanin, S. V. Morozov, L. A. Ponomarenko, R. V. Gorbachev, A. S. Mayorov,
M. I. Katsnelson, K. Watanabe, T. Taniguchi, K. S. Novoselov, L. S. Levitov, and A. K. Geim, Science {\bf 332}, 328 (2011).

\bibitem{GiantSH}
Qing-feng Sun, X.C. Xie, Phys. Rev. Lett. {\bf 104}, 066805 (2010);
D. A. Abanin, R. V. Gorbachev, K. S. Novoselov, A. K. Geim, L. S. Levitov, Phys. Rev. Lett. {\bf 107}, 096601 (2011).

\bibitem{Nikolic12}
C. -L. Chen, C. -R. Chang, B. K. Nikolic, Phys. Rev. B {\bf 85}, 155414 (2012).

\bibitem{RibonFerro}
Y.-W. Son, M. L. Cohen, and S. G. Louie, Nature (Lon-
don) {\bf 444}, 347 (2006).

\bibitem{grapheneFerroMetal1} E. W. Hill, A. K. Geim, K. Novoselov, F. Schedin, and P. Blake, 
IEEE Trans. Magn. {\bf 42}, 2694 (2006).

\bibitem{grapheneFerroMetal2} N. Tombros, C. Jozsa, M. Popinciuc, H. T. Jonkman, and 
B. J. van Wees, Nature (London) {\bf 448}, 571 (2007).

\bibitem{grapheneFerroMetal3}  C. J\'{o}zsa, M. Popinciuc, N. Tombros, H. T. Jonkman, and 
B. J. van Wees, Phys. Rev. Lett. {\bf 100}, 236603 (2008).

\bibitem{InsulFerro}
A. G. Swartz, P. M. Odenthal, Y. Hao, R. S. Ruoff, and R. K. Kawakami, ACS Nano {\bf 6}, 
10063 (2012).

\bibitem{Chud09}
E. M. Chudnovsky, Phys. Rev. B {\bf 80}, 153105 (2009).

\bibitem{GrapheneSH01}
C. Weeks, J. Hu, J. Alicea, M. Franz, and R. Wu, Phys. Rev. X {\bf 1}, 021001 (2011).

\bibitem{GrapheneSH02}
A. H. Castro Neto and F. Guinea, Phys. Rev. Lett. {\bf 103}, 026804 (2009).

\bibitem{GrapheneSH03}
C. Ertler, S. Konschuh, M. Gmitra, and J. Fabian, Phys. Rev. B {\bf 80}, 041405(R) (2009).

\bibitem{GrapheneSH1}
H. Zhang, C. Lazo, S. Blugel, S. Heinze, and Y. Mokrousov, Phys. Rev. Lett. {\bf 108}, 056802 (2012).

\bibitem{GrapheneSH2}
P. Ghaemi, S. Gopalakrishnan, and T. L. Hughes, Phys. Rev. B {\bf 86}, 201406 (2012).

\bibitem{GrapheneSH3}
N. A. Sinitsyn, J. E. Hill, Hongki Min, Jairo Sinova, A. H. MacDonald, Phys. Rev. Lett. {\bf 97}, 106804 (2006).

\bibitem{KaneMele}
C. L. Kane and E. J. Mele, Phys. Rev. Lett. \textbf{95}, 226801 (2005).

\bibitem{SpintronicsRev1}
I. Zutic, J. Fabian, S. Das Sarma, Rev. Mod. Phys. {\bf 76}, 323 (2004).

\bibitem{SpintronicsRev2}
Y. Tserkovnyak, S. Brataas, G. E. W. Bauer, B. I. Halperin,  Rev. Mod. Phys. {\bf 77}, 1375 (2005).

\bibitem{AHERev1}
E. Hall, Phil. Mag. {\bf 12}, 157(1881);.

\bibitem{AHERev2}
 N. Nagaosa, J. Sinova, S. Onoda, A. H. MacDonald, and N. P. Ong, 
Rev. Mod. Phys. {\bf 82}, 1539 (2010).

\bibitem{AHEIn}
N. Nagaosa, J. Phys. Soc. Jpn. {\bf 75}, 042001 (2006).

\bibitem{AHEEx}
J. Smit, Physica {\bf 24}, 39 (1958); L. Berger, Phys. Rev. B {\bf 2}, 4559 (1970).
\end{thebibliography}
\end{document}